\documentclass{article}
\usepackage[english]{babel}
\usepackage[utf8]{inputenc}
\usepackage{johd}
\usepackage{endnotes}

\title{A Brief Discussion on the Philosophical Principles and Development Directions of Data Circulation}

\author{Zhi Li$^{a}$, Lei Zhang$^{b}$, Junyi Xin$^{c}$, Jianfei He$^{d}$,  Yan Li$^{a}$, Zhenjun Ma$^{a}$, Qi Sun\textsuperscript{$\ast,$}$^{a}$ \\
        \small $^{a}$Hangzhou Nuowei Information Technology, Hangzhou, China\\
        \small $^{b}$Nanhulab \\
        \small $^{c}$School of Information Engineering, Hangzhou Medical College, Hangzhou, Zhejiang, China\\
        \small $^{d}$City University of Hong Kong
        \\\\
        \small $^{*}$Corresponding author: Qi Sun \tt{sunq0810@gmail.com} \\
}
\date{}

\begin{document}
\maketitle
\begin{abstract} 
\noindent The data circulation is a complex scenario involving a large number of participants and different types of requirements, which not only has to comply with the laws and regulations, but also faces multiple challenges  in technical and business areas. In order to systematically and comprehensively address these issues, it is essential to have a comprehensive and profound understanding of 'data circulation'. 

The traditional analysis method tends to proceed based on the traditional circulation model of commodities, that is, tangible objects, which has some defects and shortcomings, and tends to be a formalized approach, which is faced numerous challenges in practice. This paper analyzes the circulation of data with a philosophical approach, obtains the new explication of data and executing entity, and provides a new definition of the concepts of data utilization and data key stakeholders (objects). At the same time, it puts forward the idea of ``data alienation'', and constructs a new interpretive framework of ``data circulation''. 

Based on the framework of this interpretation, it is clearly proposed that ``data alienation'' is the core of ``data circulation'', benefit distribution is the driving force, and legal compliance is the foundation, and further discussed the three modes of ``data circulation''. It further discusses the three modes of ``data circulation''. It is pointed out that ``data circulation'' is different from traditional ``commodity circulation''. To achieve ``data circulation'',a comprehensive information infrastructure needs to be established. from a theoretical point of view, it lays a solid foundation for the development of ``data circulation''.
\end{abstract}

\noindent\\



\section{Introduction}
Exploring ``data circulation'' from a philosophical perspective is a novel interdisciplinary challenge that has attracted the attention of philosophers, computer scientists, information scientists, cognitive scientists and ethicists, lawmakers, and even researchers in different fields such as policy research. Philosophy, as the science of sciences, can enhance understanding and cognition, and the philosophical speculation can help us to construct an analytical and speculative space with ``data circulation'', allowing for meaningful interpretation.

The emergence of new technologies such as privacy-preserving confidential computation, which are contrary to the perception of common experience, have led to significant ``confusion'' among those involved. How can data be ``available but not visible?'\footnote{  `handles learning from distributed sources without sharing raw data.'' This may be the first document to propose the idea of ``Data Availability but Invisibility'', and gives a concrete technical implementation plan.} 'What technology is being adopted?'\endnote{Shuang Wang, Xiaoqian Jiang, Yuan Wu, Lijuan Cui, Samuel Cheng, Lucila, `Ohno-Machado, `handles learning from distributed sources without sharing raw data.'' Distributed privacy-preserving online model learning: , EXpectation Propagation LOgistic REgRession (EXPLORER): distributed privacy-preserving online model learning, Journal of Biomedical Informatics, Volume 46, Issue 3, 2013, Pages 480-496,ISSN 1532-0464, \url{https://doi.org/10.1016/j.jbi.2013.03.008}. (\url{http://www.sciencedirect.com/science/article/pii/S1532046413000427}). \\ 
} How does this affect the way data is processed? How did it overturn conventional views? In this paper,  we start with exploring the underlying connotation of the concept of data foundation at a deeper level, utilize philosophical analysis methodology to analyze and introduce several key concepts of data circulation, and propose the construction of an information infrastructure centered on a ``secure computing space for data value,'' in an attempt to achieve the goal of ``data circulation'' through the development of a ``secure computing space for data value''. It also proposes the construction of an information infrastructure centered on the ``secure computing space for data value'' in an attempt to theoretically put forward a framework for solving various technical and operational problems in the marketization of data factors by taking ``data circulation'' as an entry point.

Compared to traditional things, data possess unique characteristics: data often do not exist alone, need to dependence on a physical form of carrier, medium. In addition, Data can be easily copied,easy to be tampered with, the occurrence of abnormalities in the existence of the difficulty to be detected and verified and other issues. These phenomena are called ``vulnerability'' in this paper\footnote{Vulnerability: English word ``vulnerability'', susceptibility to harm, difficulty in defending against possible attacks and damage. Vulnerability of data draws on concepts related to data security in the industry and also includes issues or concepts such as susceptibility to illegal copying, tampering, and difficulty in sensing unauthorized access.}.

Some fundamental concepts related to data are controversial and poorly understood, especially when the data involves personal information, and there are deep and easily overlooked detailed concepts that need to be clarified in order to gain a comprehensive and profound understanding and insightful perceptions.

In recent years, the enactment of the data market bill represented by the European Union at the international level and the concept of data factor marketization at the domestic level have shown that the prevailing view of the society has gained sufficient knowledge about the increasingly important role of data in the society and economy. However, the concepts related to data are very complicated, such as data factors, data transaction, data exchange, data value and data circulation, etc., and the definition and understanding of data are not consistent. Among them, ``data circulation'' is a concept that has been widely mentioned recently and involves a series of details that have been neglected in the past. These relatively new concepts involve new technologies. How to recognize and gain insights into the essential problems and propose innovative solutions requires a philosophical thinking and analysis that goes back to the origin of data. If we can apply philosophical and speculative methods for thinking and analysis, we can achieve better understanding, which can then guide specific technical and business management tasks.

The concept of data circulation is indeed quite new and encompasses a wide range of concepts and meanings, which makes it a challenge for most non-computing professions, especially those with a background in social sciences, to understand data circulation. Even the computer industry people have have significantly different understanding of this. The purpose of this paper is to let readers, particularly non-data industry readers, have a basic knowledge of data, especially data circulation and related concepts from a non-specialized technical point of view in a relatively short time, and try to recognize the basic vein of data circulation from the complicated world, And this provides a starting point for forming a basic understanding of the entire data industry, thereby enabling the selection of a better data  industry development trajectory and path.

On the other hand, objectively speaking, ``data'' and ``data circulation'' are relatively ``big'' topics, in order to avoid generalization, and to support and guide specific applications in concrete practice.  For a more focused discussion, this paper focuses on the more specific and practical topics of ``data circulation''. From this point of view, we take data as the research object, take the process of data circulation as the core, We analyze it, clarify its mechanism, and build a cognitive framework. In this paper, we will clarify the challenges faced by ``data circulation", and point out that ``data circulation'' is a scenario involving multiple stakeholders, and will also face various technical and business challenges. ``Data circulation does not exist independently, and to meet specific business requirements, it needs to be supported by sufficiently complete technical means.

Finally, this paper proposes a feasible overall program, we call it,''Security Computing Space for Data Value''. By constructing a data circulation framework from the perspective of technology and business integration, it is able to solve various problems in data circulation in a relatively complete manner, especially the problem of safe circulation under the premise of compliance, and lay the foundation for the full development of data circulation.

This series of articles will also be a preliminary introduction to the logic, business, and technical architecture of the ``security computing space for data value", the details of the specific content will discuss in the other articles in this series.

Please note that this paper presents a number of more novel concepts or differences from current understandings and definitions:
\begin{itemize}
    \item Data: there are some differences in the concepts presented in this paper compared to the definitions presented in other literature. In the scenario identified in this paper, data are the results of a metric comparison.
    \item Data utilization: Data being used is the process of being used to compute, the process of deriving value through computation.
    \item Data alienation: the essence of data utilization, this paper proposes that the essence of data utilization is the data alienation , which centers on performing calculations and generating new data. The original key stakeholders or key stakeholders (see the concept of the fifth bullet point) to which a particular data is bound and the associative relationship between the data are stripped away. The newly generated data as a result of the calculation has an essential change in the meaning of the original data. In this process, new key stakeholders or associative relationships are formed, which is the alienation of the data. Furthermore,, the alienation of the data is not only a method that ensures data safety during data utilization but also a compliant approach for data usage.
    \item executing entity: The entity responsible for the operation of the data, such as a person, a machine, an organization, a legal person, etc. From the point of view of legal responsibility, the executing entity should assume the corresponding duties and obligations and enjoy its powers because it is able to operate on the data.
    \item Data key stakeholder (stakeholder entities): Similar to concept in the project management, the concept of a stakeholder. Project stakeholders are individuals, groups, or organizations that can influence project decisions, activities, or outcomes, as well as individuals, groups, or organizations that will be affected by project decisions, activities, or outcomes. Data stakeholders, on the other hand, are the people, organizations, institutions, entities, etc. to which specific data are related. Data with which the most direct and close direct correlation relationship, called the core stakeholder, core stakeholder.
\end{itemize}
These concepts, and related concepts, will be further discussed and clarified in following sections of this paper. Due to space limitations, much of the content in this paper consists of direct quotations or listed conclusions that are not described and discussed in depth. Please refer to the professional literature or explore other articles for related content.

\subsection{Summary}
The concept of ``data circulation'' is an interesting one, and similar or related to it are the concepts of ``data trading", ``data exchange", etc., which are understood and interpreted in different ways due to different backgrounds, experiences, purposes, etc. in the industry. Due to differences in industry background, experience, purpose, etc., there are different understandings and interpretations of this concept. First of all, it is necessary to clarify what is ``data", what is ``circulation", what is ``data circulation", as well as the related basic concepts, these seemingly simple concepts, which are less emphasized, are also more obscure and underlying. These seemingly simple concepts, commonly ignored in the past, which are relatively unappreciated and obscure, are complex and controversial when examined in detail. Clarifying these concepts will help us to further understand the underlying logic before exploring the concept of data circulation. Therefore, we first need to explore what is data, what is the source of data, what is the use of data, and how to use the data are a few important issues. After clarifying ``data circulation", we will compare and analyze the concepts of ``data transaction'' and ``data exchange''.

This section focuses on the preliminary discussion of the philosophical meaning of ``data circulation", the above mentioned many data circulation scenarios, here we do a rough analysis from the philosophical point of view. First, let’s explain and examine, from the perspective of philosophical analysis, what data is and what data circulation entails. We will explore where it occurs, its processes, and its key characteristics and attributes.

\section{What's the Data?}

First of all, what is data? On the surface, data is nothing more than records in the form of text symbols and numerical symbols. Wikipedia explains it like this:

Data, also known as information, are numerical characteristics or information obtained through observation. More specifically, data is a set of qualitative or quantitative variables about one or more persons or objects. Data can be a pile of magazines, a stack of newspapers, minutes of a meeting, or a patient's medical record.

Although the terms ``data'' and ``information'' are often used interchangeably, their meanings differ significantly. In some popular publications, ``data'' becomes ``information'' when it is examined in context or analyzed; in general, data is processed and called information, and the messages analyzed from this information are called knowledge. Generally speaking, data is called information after processing, and the message analyzed from this information is called knowledge, which is then gradually formed into wisdom through continuous action and verification. After the rise of big data, the discipline of data science has become very important. However, in the discussion of academic entities, data is only a ``unit of information''. In the economic field, ``information'' gives rise to the information economy, while ``data'' gives rise to the data economy.

The above is an explanation from Wikipedia.

Baidu Encyclopedia explains Data is a symbol that records and can be identified for objective events, a physical symbol or a combination of these physical symbols that records the nature, state, and interrelationships of objective things. It is recognizable, abstract symbols.

It refers not only to numbers in the narrow sense, but also combinations of words, letters, numerical symbols, graphics, images, video, audio, etc. with certain meanings, as well as abstract representations of the attributes, quantities, locations of objective things and their interrelationships. For example, ``0, 1, 2...", ``cloudy, rainy, falling, temperature", ``students' file records, transportation of goods", etc. are all data Data is processed to become information. Data becomes information when it is processed.

In computer science, data is a general term for all symbols that can be entered into a computer and processed by a computer program, and is a generic term for numbers, letters, symbols, and analog quantities that are used to input into an electronic computer for processing and that have a certain meaning. The objects stored and processed by computers are very extensive, and the data representing these objects are becoming increasingly complex as a result.

The above is the explanation in Baidu's encyclopedia.

It should be recognized that Wikipedia and Baidu's explanations are not very scientifically rigorous and are for informational purposes only. No one has attempted to define and interpret data from a philosophical point of view. In the book ``Philosophy of Information'' by the philosopher Luciano Floridi\endnote{Floridi, Luciano, "The Philosophy of Information", Oxford University Press Published: 27 January 2011
 ISBN: 9780199232383 by Luciano Floridi }, it is argued that data are basically equivalent to symbols, while information is defined as ``data'' plus ``meaning''. From this point of view, data and meaning are independent of each other. This paper takes a different view and argues in this paper that it is the phenomenon rather than the essence of data.

Data is a relatively abstract concept, but its essence may be very clear. According to the authors, data can be thought of as the result of comparing some aspect of something (i.e. a stakeholder entities) with a corresponding benchmark, and this result is stored and used in some symbolic form and in some specific way. Further, from a practical point of view, this result should be expressed in a systematic symbol system that can be processed by a computer, for example, using linguistic and numerical figures, which can be conveniently stored and transmitted by a computer system, and computed. From a philosophical point of view, data can be considered as the ``alienation'' of the result of the comparison, as well as the form and carrier of the ``alienation"\footnote{``Alienation'' is a very basic philosophical point of view, especially the important concepts and ideas in Marxism, but also more profound. It is also the core idea of this paper, please refer to https://baike.baidu.com/item/\%E5\%BC\%82\%E5\%8C\%96/5807932?fr=ge\_ala for preliminary understanding, and please refer to professional philosophy books for in-depth understanding, ``data alienation'' is an important concept in this paper. In this paper, ``data alienation'' is an important concept, with the help of the concept of ``alienation'' to define the data computing can be more accurate insight into the data process and cognitive grasp.}.

Of course, what is not stored in the computer system is also data. However, processing the circulation of such data is too costly and cost-inefficient. In principle, this type of data also applies to what is discussed in the text. From the broad definition, also suitable for 70,000 years ago primitive man recorded in the stone scratches\footnote{In 2002, in the Blombos Cave in South Africa, archaeologists discovered stones with grids or cross-hatched patterns dating back 70,000 years. While some studies have attributed this discovery to ``primitive art", authors have argued that it is more likely to be a record, based on the theory of ``knotting''.}, which can also be attributed to data.

From this point of view, data is not ``a priori"\footnote{Here draws on the concepts of a priori and a posteriori knowledge, a priori and a posteriori are important and fundamental philosophical concepts, a detailed discussion of which can be found in specialized philosophy books, and from a superficial point of view, can be superficially understood in this way: knowledge that relies on experience, facts, and evidence is a posteriori, and knowledge that does not, and that relies only on rationality, is a priori, e.g., mathematics.}, but symbols are a priori, It is the relationship between the symbols and the key stakeholders that makes the symbols become data for those key stakeholders. The essence of data is the result of measurement and metrics of specific attributes, states, and processes of key stakeholders. The objective effect is that digital symbols are associated with key stakeholders and stakeholders.

Put another way, data is not self-interpreting, and the connotative interpretation of data involves issues of semantics and context, especially the associative relationships between the stakeholders and objects associated with it, and the time-space in which it is located.

Of course, the concept of data can be further explored and analyzed from multiple perspectives in the context of pure mathematics, informatics, logic, cognitive science, legal science, and so on, but it is beyond the scope of this paper. The main concern of this paper is the scope of the theme of ``data circulation", which is a more pragmatic and specific and narrower scope of the link.

So how do associative relationships between symbols and stakeholders (or stakeholder entities) arise? This brings us to what comes next: the sources and generation of data.

\section{Sources and Generation of Data}

In the previous section, we obtained an abstract notion of data, i.e., data can be thought of as a comparative result about some aspect of something and a corresponding datum, and this result is usually saved stored, transmitted and used in a particular way, in some symbolic form that minimizes the total cost.

Further analysis based on ontological approach\footnote{Ontology: There is no consistent definition about ontology, here refers to the ontology of some of the analytical methods and analytical ideas, ontology is a deep and tedious philosophical concepts, detailed discussion far beyond the scope of this article, interested readers can refer to the professional philosophy books.}, further analysis, to carry out metrics is an action, this action is not generated out of thin air, but there must be a role in the execution of the action, if the ``action'' is regarded as a predicate, there must be a ``entity'' and the ``predicate'' directly related to the ``predicate''. If ``action'' is regarded as a predicate, then there must be a ``entity'' that is directly related to this ``predicate''. In this paper, the role of executing the action is called ``executing entity", this role is responsible for executing the action ``comparison", and the results obtained are data. The role of the executing entity is very critical. It is the initiator and executor of the action in question, and only it has the ability to ``act"\footnote{Energetics, also a basic philosophical concept, the emphasis here is on its initiative, the ability to initiate actions to perform functions. Please refer to specialized philosophical books for detailed concepts.}, especially the initiative, the ability to initiate the execution of the action, which is not the case for a stakeholder or a stakeholder.

Data is about a specific thing, which is referred to in this paper as a key stakeholder/key stakeholders entities. For example, this thing can be a person, an object, and an aspect, such as a person's height, an area of land, a process, or a more complex multidimensional measure. The key stakeholder/key stakeholders entities is the most critical attribute of the data, and the key stakeholder/key stakeholders entities to which the digital symbols are connected through metric measurements as described earlier, and is a key factor in the various issues facing the flow of the data, especially those related to the law. The form and content of this connectivity through measurement are also very complex, but this factor has little impact on the initial analysis of ``data circulation'' and is therefore beyond the scope of this paper.

The resultant expression of a metric is related to the cognitive structure of humans nervous system\endnote{ Eric Kandel, James Schwartz, Thomas  Jessell, 
 "Principles of Neural Science"   McGraw-Hill Medical ISBN 9780838577011 }. The nervous system' of human beings always gain understanding by comparing against a certain foundation, baseline, or benchmark.\footnote{This is based on the ability of biological tissues and organs to be fit for survival, developed through a long evolution, please refer to the specialized information for more details.}, for example height in centimeters and land area in square meters. The structure of such comparisons can be thought of as the result of the collection of measurement metrics, For example, one person's height could be 1.68 meters, while the area of a plot of land could be 30 square meters.

Note the act of comparison here is performed by an responsive executing Entity,which could be an abstract concept, the thing itself, or something else. The result of the comparison is obtained from the executing Entity, although it is related to the thing being measured (that is, stakeholder/stakeholder entities). The entity performs the behavioral action of measuring and comparing, and the result is usually represented in the form of symbols, so that the symbols are associated with key stakeholders and key stakeholders, and the symbols that are associated become ``data''. For example, if a blood type test is performed on a person, the entity of the test is a medical device, and the result is related to that person, and the content of the measurement may be private, but the result data is obtained from the medical device, not the person. That is to say, the result obtained from this measurement is the establishment of a relationship, and data that is not associated with stakeholder/stakeholder entities is defined in the scope of this paper as mere symbols. In the view of this paper, we believe that data cannot stand alone, while symbols can stand alone.

Measurement results must be recorded and presented in some form, usually a symbolic system of expression, e.g., numerical, verbal, and stored in an appropriate manner for use. In today's society, this usually means that they are stored in a way that can be automated by a computer and processed for use.

For example, if I, as an individual, buy a book from Taobao, it can be interpreted that Taobao is the executing Entity that measures me as an individual, and records a piece of data, ``purchased a book''. It is also possible to understand that the executing Entity is me as an individual who has measured Taobao as a thing and recorded a piece of data ``a book was sold''. The corresponding notation for this data may be in the form of a computer statement, or it may be in the form of a sentence in Chinese or some other natural language.

\subsection{Vulnerability of Data}
Data employs symbolic writing as a carrier, and symbolic writing also requires a medium as a carrier. Before the invention of computer technology, a lot of data is also through the symbolic way carved in stone, written on paper, to a large extent, the use of stone and paper is difficult to copy the characteristics. Stone paper is tangible and different between individual objects can be done easily distinguishable, engraved stone, written on the word of paper or printed paper, although the word is easy to copy, but its dependence on the stone paper is easy to distinguish, so copying and tampering with the existence of considerable difficulty and cost (refer to the characteristics of banknotes).

The situation changes significantly when the carrier of the symbols takes the form of a computerized technical device. Computer as the carrier of the mode of natural existence of easy to copy, tamper with the characteristics of the principle that the data is very ``fragile", that is, easy to copy, easy to tamper with, not easy to trace, which is the characteristics of the computer as a carrier. Especially the new generation of computer technology, communication technology popularization, the cost of data dissemination in most cases can be negligible. The problem of insufficient and defective innate conditions underlying the right to data has been further exposed. After an item is used, its physical state changes more or less, while data does not, and usually data does not change after use\footnote{The use of ``security computing space for data value'' technology to realize the information infrastructure can be realized under certain conditions,  after usage, data may be changed or ``destroyed", which is one of the characteristics of the ``security computing space for data value''.}, while generating new data. This natural attribute of data is the fundamental reason for the difference between traditional items, the nature of data is non-object, according to the traditional object as the object of the management of the way there are many inherent, fundamental flaws. For example: traditional stock trading, futures spot trading has a set of basic support structure. The core of its foundation is the entity matter of the replicability is not strong. The extreme reproducibility of data, ``vulnerability'' makes the establishment of a data exchange in accordance with the model of the stock exchange does not have the practical feasibility, which is also a lot of traditional goods in the form of commodities, regulatory systems, circulation methods are not quite applicable to the ``data'' reasons. This is also the reason why many traditional commodity forms, regulatory regimes, and circulation methods are less suitable for ``data", and one of the factors to be considered and overcome in a data circulation approach.\footnote{Some scholars have proposed the use of intellectual property protection methods for data management, such as patents, trademarks management program, this paper argues that this approach, intellectual property regulation program for data, the granularity is too coarse, the cost of management costs is too large, but also can not meet the compliance requirements of laws and regulations, and is not safe, the practice is not very feasible.}

\subsection{Key Stakeholders}
Refers to the object being measured in the data generation process. key stakeholder, specifically refers to the result of the measurement and the object being measured with personality. Because key stakeholders may be directly or indirectly involved in specific personalities, they may be entity to many laws and regulations such as privacy protection. Also because of their person-hood, they have a range of potential claims to the data associated with them. Also because the use of such data may affect or harm their rights and interests, they require adequate protection from the parties involved, and they may also require participation in benefit also because of the potential gains from the use of such data by those parties. If the object being measured is an object, there may be other players that make demands or ask to participate in benefit distribution because of legal factors such as the right of attribution of the object. In reality, there are more complex scenarios with multiple key stakeholders, which are not addressed in this paper to simplify the discussion.

\subsection{executing Entity}
Data does not emerge out of nowhere, data comes from a specific action, that is, the comparative behavior, that is, the behavior of measurement, metrics and collection and analysis, the behavioral action is usually carried out by the entity of the implementation of\footnote{The executing Entity, in practice, and data related to the right of attribution, the right to confirm, property rights and other concepts is difficult to clearly define a consistent and common concept. In this paper, we propose a concept of data-related executing Entity. The executing Entity is a very easy and operational as well as practical application of the concept, there are significant differences between the two, in a particular scenario can be mapped and associated.}, and the result of the implementation of the behavior is the generation of data. The executing Entity is dynamic and active, and the data is the result and the purpose; while the core data stakeholders (objects) are passive. In other words, the actor who performs the comparative behavior at the stage of data source generation, is the executing Entity. The executing Entity has the power to manipulate the data, and the generation of data is also a type and link of data manipulation.

The executing Entity is related to the thing that is the object of execution and the result of execution, and it is the first acquirer of data in terms of chronological order. Therefore, the executing Entity bears the initial responsibility for the legality and accuracy of the data, etc., and is also the object of the legal rules that are required to be imposed.

The object of comparison is the key stakeholder or key stakeholder entities of the data, and there is also a natural connection between the data and the key stakeholder or key stakeholder entities, which is ``a priori", ``exposed'' or ``acquired", ``known'' and ``discovered'' through metric measurements. This natural connection is ``a priori", and is ``exposed'' or ``acquired", ``known'' and ``discovered'' through metric measurements. This natural connection is also one of the relational bases of the right to data and one of the bases of the legal rules that ``require''. This natural connection is also one of the relational bases of the right to data and one of the bases of the ``requirements'' of legal rules.

When the core stakeholder of the data is a ``person", be it an individual or a group, there are a series of strict legal requirements that need to be met for the protection of the individual group. Only the executing Entity can fulfill this requirement, that is, the executing Entity naturally assumes the relevant obligations.

Further analysis shows that there is an interesting fact that the executing Entity is supposed to know about the data, whereas the key stakeholders related to the data are not necessarily aware of it, and even if they are aware of it, they get it from the executing Entity. Only when the executing entity and the relevant core stakeholder overlap, the core stakeholder is able to know the data directly.

\subsection{Boundaries of the executing entity}
Technically, from a business logic transaction perspective, an executing entity can be an independently constituted entity\footnote{Entity: ``Entity'' is a very important and fundamental concept in philosophy, as well as an esoteric concept that different schools of philosophy and philosophers have their own understandings and interpretations of, and actively use. Usually in ontology, ``entity'' refers to something that ``exists'' and may or may not have a physical form. In this paper, the concept refers to an organization that is logically and logically identical in a legal sense, e.g., a company, a hospital, a system, etc. In this paper, the concept of ``entity'' is used to refer to an entity that has a physical form. In this paper, the implementation of the entity and the entity of the two concepts are logical concepts, there is a certain degree of overlap, the implementation of the entity is mainly biased technical point of view, the ability to complete a particular task of the unit, the entity is biased towards the legal and management point of view, to assume a particular responsibility, especially specific legal duties of the unit. From a practical point of view, it is common for an entity to include one or more executing agents.}, or it can be a combination of multiple executing entity to form an independent entity. When it is an independent entity, from the legal point of view, it is the executing entity who independently assumes the rights and obligations of data access. When a combination of executing entity constitutes a separate entity, the rights and obligations to access data are shared with multiple executing entity. The executing entity has substantive ability, but its ability and responsibility also has a boundary, the so-called boundary refers to the scope of the ability of the executing entity's responsibility and right, not only from the technical and operational point of view, but also from the point of view of the legal point of view or the division of responsibilities of internal management to confirm the scope. Boundaries are recognized by specific judgmental criteria, which are used to confirm whether a certain part is attributed to a specific executive entity

This criterion can be substantively confirmed through technical means, or formally confirmed in written form, or a combination of the two, has the force of law, i.e., it is only within the boundary that the entity can operate on the particular data and assume duties and obligations. This scope is also the scope of action of the executing entity with respect to his or her data, and he or she is burdened with the corresponding duties within the boundary. Crossing this boundary would result in the loss of corresponding rights and obligations. This is also the fundamental principle behind the concept of "raw data not leaving the domain".

Within the framework established in this paper. the executing entity is a very key role, is the role of entityive initiative to perform the operation, but also the role of the legal obligations passed from the entity to enjoy the rights and interests of the role of the core concepts in the interpretation of the ``data circulation'' framework in the data acquisition, data utilization, data destruction and other aspects of each scenario play a key role, and can establish a corresponding relationship with the entities in actual business operations. and will be described in detail. in the following paper.

\section{The \textbf{Role} of Data}

What is the role of data? This is a question that goes without saying for modern society, especially as technology represented by large AI models continues to make breakthroughs at a visible pace. The focus of this paper is how data is used, especially compliant cross-domain use. Focusing on the principles of data utilization, and the issues related to the changing benefits of use, this paper explores the key issues in data circulation from the purpose of use.

The purpose of data utilization is to obtain a benefit. The use of data is the manipulation of data, which requires an executing entity to perform the manipulation, and the acquisition of benefits through the executing entity. The executing entity obtains benefits from the use of the data, which should be distributed fairly and equitably among the relevant parties and in accordance with the relevant laws and regulations.

\subsection{The Nexus between Data Utilization and executing entity}

“Data utilization refers to employing data to accomplish real-world tasks. Data utilization manifests the practical value of data. The concepts of data resources, data assets and data factors are based on the use value of data. Unlike data reproduction, data utilization does not aim at producing new data, but at accomplishing a realistic task. It is important to note that if the purpose of data utilization is to accomplish some task in cyberspace, then the process will be data reproduction. `` --Data Autonomy\endnote{https://developer.aliyun.com/ebook/download/7467}

Once data has been generated and acquired, it usually needs to be stored, transmitted and used. Usually, after the data has been acquired, it often needs to be processed to meet certain requirements, transmitted and stored, which is also known as ``data cleansing'' or ``data governance''. ``The data can be used only after the operations corresponding to these concepts.

In the view of this paper, data utilization is a process that is highly purposive, and is essentially a process of ``alienation'' aimed at obtaining a computational result to be presented to the appropriate entity. Only a specific computational manipulation of data that produces a result, which is a substantially new result, can be characterized as ``data utilization''.
Typically, the manipulation of data can be categorized as data utilization. However, strictly speaking, the transmission and storage of data does not constitute the use of data, even the traditional ``data cleansing'' or ``data governance", the data does not have a relationship with its key stakeholders and stakeholders cut, change, the above behavior is just to The above behaviors are just for the better use of data. Data has not been ``alienated", therefore, according to the viewpoints and scope of this paper, this phenomenon does not belong to ``data utilization", but is only one of the preparatory work related to data utilization. This is also an important difference between this paper and the traditional views.

The utilization of data also requires an executing entity, which calculates the data and gets the result, and this result is also data. This data, there are also key stakeholders, stakeholders. But it is not exactly the same as the original data. This executive entity, may be the same entity of data acquisition, can also be other entities.

What is the philosophical definition of ``data utilization''? In this paper, we propose that data utilization can be regarded as the cutting of the relationship between data and its specific stakeholders and objects. There is a corresponding relationship with the concept of data generation or acquisition, in which data generation and acquisition is a connection with a specific stakeholder or stakeholder, whereas data utilization is a cutting of the original connection, where the symbols that represent the data are changed, and the data are divorced from the relevant key stakeholder or key stakeholder entities, and connected to a new key stakeholder or key stakeholder entities, resulting in the ``alienation'' of data. ``Alienation'' is the real use. Alienation here can be understood as the generation of new data, and relative to the original data, the symbolic text representing the data has changed, and the relationship has changed, and the data's key stakeholders have changed, and these changes may occur simultaneously or separately.

Often many scenarios in the ``data utilization", the concept of this term, but did not carry out an in-depth study, in order to clarify the meaning of which we further explore, in many cases, the so-called ``data utilization'' is only a copy of data or data diffusion, or simply processing the actual implementation of the entity with the ability to operate. In many cases, the so-called ``use of data'' is merely the copying or diffusion of data, or simply the processing of data, with the addition of entities who actually have the ability to perform the actual operation. However, there is no ``alienation", and from the perspective of this paper, these are not ``data utilization", and the act of adding an executing entity still poses security compliance risks and challenges to both the process and the entity from a legal perspective. One scenario that needs to be clarified in particular is the transfer of ownership of data sources, which is also not a data utilization. To illustrate: 
Suppose a coal mine produces a steady stream of coal, which was originally owned by Bob, but it is now bought by John, so the ownership of the coal mine changed. The attribution of ownership of the data is also similar with it. There is no data utilization in this case either, just a change. a change of attribution of the source of the production of the item. The attribution of the data source can also have a similar treatment. There is no data utilization in this case either, just a change in the executing entity in the data generation phase. For the data that have been generated, even if the above changes have occurred, they are only based on the technical behavior of data copying and moving, but may only increase the number of executive entities who have access to the processing of the data, resulting in the complication of the corresponding responsibilities and rights of the data, which is not a circulation of the data, but at most a ``data dissemination''.

Further, logically analyzing the data usage scenarios, we find that there are three executing entity:
\begin{itemize}
    \item executing entity that perform utilization data and execute calculations
    \item Acceptance of the entity of the implementation of the results
    \item Implementing entities using data sources
\end{itemize}

Each of these three entities of execution has a corresponding entity, which may be the same or different. In the first case, if they are all the same entity, it is also the case that the whole process of use is completed under the same entity (the same technical device, the same institution, and articularly the same entity in a legal context), i.e., it means that the whole process of use of the data is completed under the same institutional organization. In the other case, the three entities mentioned above are not under the same institutional organization. That is to say, the data is not completed within the same organization, these two cases need to obey the laws and regulations, there is a significant difference, which also leads to the question of whether the use of data is legally compliant. In addition, there are also differences in technical implementation, distribution of benefits and other related issues. The first situation from the perspective of this paper does not belong to the category of data circulation, the second situation belongs to the category of data circulation.

It should be noted that according to the characteristics of the data, the original data does not change when the data is used. It can still be used again, unlike the use of other items, there is wear and tear or other physical characteristics of the items change. The use of data does not exist in the loss, there is no change, but only in the time and space has changed, the effectiveness of its use may change\footnote{A good technical solution for data circulation should be able to realize that data cannot be reused after they have been used, i.e., to overcome the ``vulnerability'' of data. Through good technical solutions, such as the ``secure computing space for data value", it is possible to realize that in the whole process of ``data circulation", the executing entity who uses the data, after using the data in accordance with the algorithm of the realization of the agreement, will not be able to use the data again.}.

Data may go through several processes from generation to use, and in terms of form and state, there may be raw data, preprocessed data, encrypted data, intermediate data, irreversible intermediate data\footnote{Raw data: preprocessed data, encrypted data , intermediate data, irreversible intermediate data, computational results, intermediate data, data using the intermediate product of the computational process, may need to be passed in the computational nodes of different executing entities, if this intermediate data can not be extrapolated back to the original data, can be called ``irreversible intermediate data", this processing is technically considered to be secure, and vice versa, is not considered to be secure. If this intermediate data cannot be extrapolated back to the original data, it can be referred to as ``irreversible intermediate data", which is technically considered secure, and vice versa. The detailed discussion of this belongs to a more complex technical domain. This paper only cites technical conclusions and results.}, and computational results, and so on. Some forms of these data may need to be transferred among different executing entities, and the corresponding executing entities need to manage and process these forms of data in a secure and compliant manner, especially to overcome the ``vulnerability'' of the data and to ensure compliance and security, which is the key to ensure data circulation\footnote{In practice, it involves more complex technical solutions and multiple trade-offs.}.

\subsection{Interests, Rights and Rules of Data Utilization}

Why is data utilization necessary ? The key objective of data utilization is 'to use'. And it is not about acquiring data, nor about obtaining access to the data. The purpose of data utilization is undeniably the pursuit of profit through the use of data, that is, the use of data to perform calculations, and through the results of the calculations, directly or indirectly, to obtain benefits, typically executing entity of the use of data to obtain the results of calculations, for executing entity of the results of the calculations and its related entities to obtain benefits or access to the potential benefits. It should be emphasized that the acquisition of data is not the acquisition of a benefit if the data is only stored, preserved without calculation, and not ``alienated''. In many cases, it is merely a preparatory step for the acquisition of a benefit.

First, we analyze the interests of the key stakeholders of the data (and, as the rationale also applies to key stakeholders, the relevant interests of the corresponding owners of the key stakeholders). Obviously, through the use of the data, i.e., the calculation, the interests of the stakeholders of the data involved in the calculation may be directly or indirectly harmed or have the potential to be harmed, or may be unaffected or minimally affected, acceptable, or gained, with an increase in benefits and possibilities. Changes in the interests of different data stakeholders in the same data utilization process may be different. Typically, the benefits to the performing entity who obtains the results of the use calculations will increase, or have the potential capability to increase.  (For convenience, if there is more than one entity of implementation of the results of the calculation, it is considered to be multiple use of the calculation.) The entity of implementation who performs the calculation, and also participates in it, has a secondary and subordinate role that is not considered for the time being.

Here the executive entity of the use of the results of the calculation to obtain the benefits, but also consider whether the other stakeholders of the executive entity of the interests of the change, whether it is an increase or decrease, each other is a ``zero-sum game", ``negative-sum game,'' or Is it a ``zero-sum game", ``negative-sum game", or ``positive-sum game''? Are the overall interests increasing or decreasing? What is the legal basis for this change in interest? What is the basis?

These changing of the interests raise a number of questions. The most fundamental of these is, should, can, and to what extent, and in what way, should the core stem of the data protect their interests? Does the executing entity have a right to an interest? In particular, is there a right to access the benefit when there is harm or potential capacity for harm to the stem person of the particular individual data? And is there a proactive approach to acquiring this benefit, and if there is a desire to protect, is there an appropriate way and ability to do so, and what is the cost of having protection and what is the cost of acquiring the benefit? Does the protection of a particular individual still pose a potential loss to the overall interest?

In the following discussion, in order to focus on the core key issues, this paper does not discuss the scenarios of attacks, data security, data leakage or tampering, etc. that are attributed to the execution of the entity that is illegal in the interpretive framework in this paper.

First, the executing entity, which the use the results the results of data calculations. This executing entity must be a part of a business  entity, whose  interest should increase, and if it decreases, there is no incentive to let executing entity to execute the calculation.

If the benefit to the entity of the usage of the result of the data calculation is smaller than the benefit to the stakeholders of the data, i.e. the use of the data leads to a loss of the overall social benefit, then the use of the data can be interpreted as a ``negative-sum game"\footnote{Negative-sum game: Simply put, after the game: both sides lose, both sides suffer losses, but also includes a party to obtain the benefits received less than the other party received losses, but the overall interests are damaged. Zero-sum game is that one party obtains the gain is the other party's loss, the overall interests have not changed.} and is usually not enforced or prohibited. If the increase in benefits from the implementation of the results of the data calculation is greater than the loss of benefits to the stakeholders of the data, i.e., if the use of the data contributes to an increase in the wealth of society as a whole, then the use of the data can be interpreted as a ``positive-sum game"\footnote{Positive-sum game: a game through which the interests of both parties are increased, or at least the interests of one party are increased, the interests of the other party are not jeopardized, and the interests of the whole are increased. Also called cooperative game.}, which will usually be encouraged and adopted. Of course, this increase in benefits is usually reflected in the distribution of benefits, which is dominated by the entities who implement the results of the data.

This pattern of interests emerges where the key stakeholders of the data, the executing entity of usage the data results, and the society as a whole each act to maximize their own interests. Due to the existence of such an interest distribution mechanism, there should be a trust mechanism between the executing entity with the malicious model\footnote{Malicious model: a model with malicious participants is called a malicious model. A malicious participant can perform any step exactly as the attacker wishes, a malicious participant can tamper with all inputs, outputs, and intermediate results in violation of the computational protocols for the principle of maximizing his/her own benefits, with the intention of changing altering input information, falsifying intermediate and output information, and trying to hide his/her malicious behavior from other participants, regulators. The attacker in the malicious model is active. The malicious model is a mapping of the real world, and the IT information infrastructure in which the data circulation should be able to meet the challenges of the malicious model, so that even a subjective malicious participant cannot disrupt the data computation process and achieve ``trusted results in an untrusted computing environment''.} as the starting point.

In general, the maximization of the overall social interests as a criterion for judgment\footnote{Overall social interest: This criterion seems to have a suspicion of ``collectivism''. Many viewpoints believe that, from the perspective of the individual, the starting point of the solution is: ``My data, my decision", and the public interest should not be used as a reason to force individuals to open up their data. However, from the current overall development trend, basically to respect the interests of individuals based on the overall interests of society as a starting point, involving specific data types have different balance points.} (detailed analysis is lengthy, this paper only mentions the conclusion), the result is the protection of individual privacy as the core of the promulgation of relevant regulations, (the core of the traditional data of stakeholders, stakeholders, because of the sensitive information related to the human being, so the countries have correspondingly introduced a large number of laws and regulations. Most of these laws and regulations regulate the role of data processors and other key stakeholders, and clarify the business norms, security rules and Ethical standards that they should follow. This paper will not go into depth.) The main starting point of these laws and regulations is to maximize the protection of individuals, and their purpose is to reduce or avoid the use of data in the process of the key stakeholders of the harm, or the transfer of interests to protect, interested can refer to the relevant literature, will not repeat here.

These regulations lay down the ground rules for the generation of benefits, and as a whole, thanks to compliance with the relevant laws and regulations, the foundation has been laid for the realization of the following objectives:
\begin{itemize}
    \item The implementation of the calculation results in significantly increased benefits;
    \item The increased benefits outweigh or are far greater than the potential loss of benefits to the key stakeholders of the data;
    \item The absolute value of the potential loss of benefits to the key stakeholders of the data is so small as to be negligible and acceptable.
\end{itemize}

On the other hand, there is the promotion of the widespread use of data as represented by the marketization of data factors, the Data Market Act, and so on. In a sense, it is with technological advancement and economic development that the key stakeholders of data are forced to be willing or agree to provide data related to them to participate in data circulation. It also empowers the executing agents who access and use the results of the calculations to obtain rights and benefits, and promotes the use of the data, access to the results, and access to the benefits.

There is a conflict of interest in the use of the data. In terms of objective effect, it is difficult for the key stakeholders to effectively defend their rights as individuals when they are asserting their rights. From an overall perspective, enacting relevant regulations and laws is the path with the lowest total cost and a mechanism for coordinating and redistributing interests. The protection and distribution of interests through laws and regulations promotes the use of data. (Detailed discussion is far beyond the scope of this paper)

\subsection{Benefit Distribution}
The execution entity obtains a benefit from the use of result of the calculation, does this benefit need to be distributed? Theoretically, it is possible. If this benefit can be distributed fairly and efficiently, it will greatly promote the willingness of the key stakeholders of the data and enhance the motivation of the executing entity of the data utilization. However, in the absence of corresponding technical infrastructure, due to the vulnerability of data, complexity and other factors, the realization of the cost is too high and practically impossible to achieve.

In reality, the relationship between the data, the key stakeholders of the data and the executing entity is diverse, and there is an ambiguity: that is, the rights between the data, the key stakeholders of the data, and the executing entity of the data generation. Due to the specificity of data, it has been difficult to effectively map and harmonize the rights related to the usual distribution of benefits of goods. Therefore, a benefit distribution mechanism centered on the key stakeholders is not a recommended perspective for this paper's proposal. Moreover, in reality, there will be a complex scenario of one data corresponding to multiple key stakeholders, which is very difficult to deal with.

From the view of this paper, The key stakeholders of the data, as the primary bearers of benefits in principle, have the potential ability to participate in the distribution.,But in reality, there are many detailed issues that make implementation difficult. In practice, another approach is to involve the executing entity that generates the data in the distribution. where the executing entity of the data generation participates in the distribution in practice. That is, the executing entity as the entity of the distribution of benefits. And the key stakeholders of the data need to be represented by the executing entity in order to obtain the benefits.

From a practical point of view, although there is no consistent mechanism, specific scenarios, after consultation is possible from the point of view of the interests of the key stakeholders of the data to identify the implementation of the main body of the relevant parties, such as the mechanism for the allocation of responsibilities and rights. This allocation mechanism is a manifestation of rights.

The key foundation of these mechanisms for the distribution of benefits is rights. What is ``rights", according to Baidu Encyclopedia:''A right is generally defined as a power conferred by law on a person to realize his or her interests. It corresponds to duties, one of the basic categories of jurisprudence, the core word of the concept of human rights and the key word of legal norms.''

The above is the explanation from Baidu Encyclopedia.

Rights is a very complex concept, the meaning of rights in data circulation mainly includes the meaning of the corresponding English word ``right", including is justified, reasonable, legitimate, such as survival, reproduction, education, freedom of religion and so on, and not quite include the English word `` power", that is, the meaning of power. power", which means power, but of course the concept of right also has elements based on violence, ability, or consensus. An in-depth discussion of the meaning of ``rights'' is far beyond the scope of this paper, which will only briefly touch on a few points related to the concept of data.

Intuitive instinct suggests, from the scope of data circulation, the key stakeholders of a particular data have all the rights and obligations of the particular data, which is an idea extended from ordinary objects, but this understanding from theory to practice there are huge dilemmas, only as a partial reference. In fact, due to the differences between data and traditional tangible objects, the complexity, relativity, and specificity of the data itself, there are significant differences and confusion in the use and value embodiment. For example, a certain enterprise jointly with a number of organizations used the data about you, recommended goods to you, from which you gained revenue, and you yourself also gained revenue. In this case, how to analyze the rights? How to implement the distribution of benefits?

In the context of data circulation discussed in this paper, a right can be defined restrictively as the right to receive benefits from the manipulation of data by the executing entity on the basis of the data. Of course, this right is also accompanied by obligations. and these rights and obligations are mainly based on laws and regulations or contractual agreements, with formal guarantees in place. They can also be technically restricted to achieve substantial protection.

The claims of the data key stakeholders based on their own data can be realized through the corresponding enforcement entities, which are extended to be equivalent to the rights of the data key stakeholders. By establishing such a framework, many problems in practice can be solved.

For example, common concepts such as data ownership can be subdivided with the advancement of technology and business into the right to collect data, the right to process data, the right to hold data, the right to use data, the right to operate data, the right to extend the right to summarize, the right to store, and so on. These terms, especially the concept of data ownership, usually refer to the ability to perform certain operations on data related to oneself from the point of view of the data's stakeholders, and these concepts are derived from traditional tangible objects, which are difficult to understand and operate in the face of the ``vulnerability'' of data. These concepts can be mapped to the right of the executing entity to perform operations on the data, and to the distribution of benefits between the executing entity and the data entities, and can then be applied in practice.

In this paper, the rights of the executing entity are defined according to its ability to operate on the data, thus the rights of the executing entity to operate on the data are clearer and more explicit, while the distribution of interests between the executing entity and the key stakeholders of the data is complicated. Therefore, this paper suggests avoiding the concept of data ownership in data circulation. Instead, it proposes deconstructing rights from the perspective of the key stakeholders of the data and constructing a clearly structured rights and benefits distribution framework based on the executing entity. This framework should facilitate practical implementation and maximize overall societal benefits.

\subsection{Data Dissemination and Data Utilization}
As mentioned earlier, according to the viewpoint of this paper, "data utilization'' is a computational process, the essence of which is "alienation", which is easily confused with  the action of "data dissemination" , or  similar terms like  "data transmission", "data Distribution", "data propagation", "data spreading". And what is "data dissemination". Before discussing this concept, for the convenience of readers with non-computer backgrounds, we will first introduce a review of how files are handled in the computer field. Copying, duplicating and moving files, these operations theoretically delete the original file location, and another kind of lossy copying, When a file is being copied, some differences may already exist between the original and the copy. Additionally, certain properties attached to the file, such as permissions (e.g., read-only, writable, owner), may change during the copying process. However, the essence of the file remains unchanged.

For example, if the entity of data utilization is different from the entity of data generation, then it is necessary to allow different entities to gain access to or control of the data in an appropriate manner, and this acquisition process is data dissemination. Data dissemination does not necessarily aim at the use of data, that is, data can be ``copied'' or ``transferred'' in various forms among different executing entity, i.e. ``data dissemination''. This process may also include a certain degree of data processing, computation, such as format conversion, content completion, etc., but the essence of the data has not changed. In other words, there is no so-called ``alienation'' of the data, and there is no cut-off of the stakeholders, so it is not considered to be the use of data, but only the dissemination.

How the executing entity obtains the ability to access specific data and the specific way is a very basic and important and critical operational behavior, It determines whether the executing entity operates correctly, complies with regulations, and meets the required standards. whether the data is safe, whether it is legally compliant, and whether the ``flow of data'' is compliant with the basic issues, but also an often overlooked underlying details.

"Data dissemination'' is not conceptualized in this paper as the use of data, nor as data circulation. It can be regarded as a preparatory or antecedent part of data dissemination or an ancillary and secondary part of it.

Data in the different entities of the implementation, copying and transmission, data dissemination in the data will probably be involved in some of the concept of segmentation, here will also involve some common operations, including data cleaning, data desensitization, data extraction\footnote{Operations such as data cleansing, data desensitization, and data extraction are common and universal processing methods for data processing. Data cleansing usually refers to removing or replenishing incomplete and erroneous parts of data in order to improve the quality of data. Data desensitization is the process of removing or blurring certain fields in order to protect privacy or confidential information, or to comply with specific laws and regulations, or to achieve a certain purpose. There are various specific techniques, so please refer to specialized information for details. It is important to understand that, contrary to conventional wisdom, some types of data cannot be ``desensitized''. On the other hand, data desensitization does not completely guarantee data privacy, as in the case of the HIPPA Act. However, it does provide a good basis for promoting the use of data.} and so on, these operations so that the data has undergone some changes. Essentially, if there is no calculation of the data, it is still only the dissemination of the data, this operation is not the use of the data, the data and the key stakeholders of the relationship has not been cut, therefore, the data has not been ``alienated''.

In addition, pure data encryption, in essence, is only a way to protect data, for example, data encryption is used for preservation and storage, and encryption is used for transmission, which is also a special form of data dissemination. In privacy-preserving confidential computation\endnote{privacy-preserving confidential computation: Initial Acceptance in the Privacy Confidential Computing Blueprint 2021}, encryption as a basic underlying technology, and other computer technology, cloud technology, algorithms, combined, constitute a step or link in the data circulation algorithm, At this point, the behavior or form of data, after being processed and encrypted (not just simple encryption), cannot be solely identified as data dissemination or data utilization..

The ``vulnerability'' of data makes data dissemination problematic, and uncontrolled data dissemination poses a variety of problems, as well as increasing the number of entities who may be performing ``illegal'' activities, all of which are all need to be addressed in practical technical and business solutions, relying on the perfect information infrastructure.

\subsection{Invalidation and Destruction of Data (carriers)}
The invalidation and destruction of data is a complex scenario from a practical operational point of view, but relatively simple from a principle point of view. From the philosophical principle, digital and symbolic is a priori, data is a posteriori, so the failure and destruction of data is the data and the corresponding entity of the data relationship is cut off, the data of the key stakeholders (stakeholder entities) become ``empty", that is, the meaning of the English void expressed in\footnote{Substantial cut-off can be achieved by privacy-preserving confidential computation based technology solutions, such as the Secure Computing Space for Data Value , which can realize this requirement in a more seamless way.}. Data because it is a symbolic form exists, so it is the ``a priori existence", cut off its relationship with the corresponding entity, the data loss of its significance, it is invalid. It is destroyed. To put it another way, although the symbol representing data has an a priori relationship with the corresponding core agent, when this relationship is hidden, the symbol regains its independent existence and is no longer data.

\subsection{Data-related Legislation}
As mentioned above, the use of data is for the overall interests of the greater, in fact, there may be a conflict of interest. The authors put forward the point of view is that the relevant regulations and laws always follow the principle of the overall social path of the lowest total cost, the perspective of the greatest benefit as a starting point, not only the protection of the weak, but also a mechanism for coordinating the redistribution of benefits. Because modern civilized society believes that that harm to the weak, even if a particular individual's harm is the overall society's great harm, which is also the modern society is unwilling to accept. Therefore strong protection of the privacy of the weak has been a hallmark of modern society.\footnote{In reality,  is not absolute or unconditional there is a tolerance threshold and rule-based judgment for the degree of harm or impact. There is generally higher tolerance for secondary effects, whereas tolerance is lower for direct harm and impact.} The benefits gained from data circulation must not violate the norms of society and must not come at the expense of harm to individuals, even if that harm is potential.

From the way this paper analyzes it, data is passive and the executing entity is active, so meeting such requirements of normative regulations can only require the executing entity to be responsible for compliance. Both the GDPR and China's privacy protection laws, as well as those of the United States, have corresponding requirements that can be mapped to specific entities for compliance and enforcement.

The use of data and the results of such use correspond to an executing entity and are entity to the relevant laws and business rules. At present, there are a large number of relevant laws and regulations in each country. From the perspective of this paper, due to historical reasons, some laws and regulations are not suitable, and some laws deviate from the original intention, which bring some unsatisfactory consequences in the overall view\footnote{For example, the EU's GDPR has been proven to be of limited use in protecting privacy after a long period of application and practice, but it has significantly increased the cost burden of relevant organizations, affecting economic development. The authors argue that the EU's GDPR is highly reliant on formal legal provisions, that there is room for improvement in the underlying framework, and that its role categorization doesn't make sense, and that it's too expensive to manage in practice.}, but the data circulation should still comply with the relevant legal and regulatory requirements.

\subsection{Summary}
Intuitively, data-related concepts are relatively simple, but still more complex and diverse when analyzed in depth. The above briefly discusses data-related foundational concepts by superficially using a philosophical approach and constructing a cognitive framework centered on data.

In the analysis of this paper, the entity of implementation is the core concept, and with this concept as the basis and starting point, an overall cognitive framework can be formed. In practice, there are also concepts such as ``data trading'' and ``data fusion", which can be analyzed with the help of the concepts discussed above, and the framework proposed in this paper can be corresponded and clarified.

In practice and application, data attributes are also very important topics, but the attributes of data do not have a great impact on the discussion of ``data circulation", so in this paper, we do not involve in the discussion of data attributes. We will now move on to the core topic of this paper - ``data circulation''.

\section{Preliminary Exploration of Data Circulation}
What is ``data circulation''? As a category, data circulation has its basic nature and basic laws, as well as its scope boundaries. In order to get a result that can guide the concrete application of practice, this paper chooses a smaller scope to discuss the meaning and definition of ``data circulation''.

The concepts of ``data", ``executing entity'' and ``data utilization'' are discussed above, and clarifying these concepts will help us further understand the underlying logic. Before discussing the concept of ``data circulation", let's review the relevant definitions proposed in this paper. data utilization requires an executing entity. Data acquisition requires an executing entity, and every operation involving data requires an executing entity. In order to accomplish ``data utilization", data needs to be disseminated among multiple executing agents, which involves each executing agent taking on corresponding duties and obligations.

First case: Intra-domain use. If the executing entity all belong to the same organization (entity)\footnote{A simple and preliminary method of determination: analyzing the entity of implementation of each link as belonging to the same entity in the legal sense, the entity of implementation of the use of the data, which is usually attributed to a legal entity. Often there is also a legal entity for the data being used and a legal entity for the results of the use. The same possibility exists between these three legal entities, i.e., the three legal entities may be the same entity or different entities. In the simplest case the three entities are the same entity. Simply put, if the data are collected and stored for use in the same legal entity, then the use of the data is ``self-produced''. This can then be simply ruled out as falling outside the scope of this paper, and the data can be processed in a way that complies with traditional DCMM requirements.}, for example, a hospital collects patient data and uses it only within the hospital. If the data does not cross domains, then the data utilization is ``self-produced and self-marketed", and the authors believe that such an approach is not data circulation. The first type of situation is the early data utilization scenario, the same organization within the use, according to our analysis above, the data utilization of the executive body, the executive body of the data generation and the executive body to obtain the results of the three processes, the executive body may be a different entity. If there is no change in the executing entity in the process of data utilization , then the data from the beginning to the end of a company / organization within a company / organization, or legally within the same group of companies, then it means that it is within the same organization, that is, the data did not go out of the border, and there is no data circulation\footnote{A controversial situation is that of authorization by an authority, i.e., an irregular situation, where access is authorized by special individual approval.}. In reality many early data circulation situations. Even if it involves different executing entity and is attributed to different entities, technically it is realized on the basis of the same entity organization, so there are many problems.

Second case: cross-domain use. If none of these executing entity belong to the same organization (entity), different from the earlier situation, then it belongs to the second situation, cross-domain use. With the development of economy and technology, more and more data are used in a cross-domain way, and more and more business scenarios require data from different entities, that is, the executing entity of data utilization and the entity of data generation and obtaining results may be different. The data and the executing entity are not within the same company/organization, and cross-domain situations occur. For example, if a hospital collects data and not only uses it internally, but also provides it to a pharmaceutical company for use, the phenomenon of cross-domain data occurs. According to the authors, this situation falls into the category of data circulation, which is the core of the discussion in this paper.

\subsection{Key Elements of Data Circulation}
According to the viewpoint of this paper: ``data circulation'' can be simply considered as data utilization in a cross-domain scenario. In this case, the entities of execution involved are different entities, especially from the legal point of view, these different entities of execution are attributed to different legal entities, there are clear boundaries between the entities, with their respective competencies, rights and obligations under laws and regulations, and the entities of execution attributed to these entities have their own clear control boundaries.

In simple terms from the data point of view, it can be divided into three environmental scenarios, the environmental scenario where data is generated and collected, the environmental scenario where data is used to perform calculations, and the environmental scenario where the results of the calculations are used.

\begin{figure}
    \centering
    \includegraphics[width=0.9\linewidth]{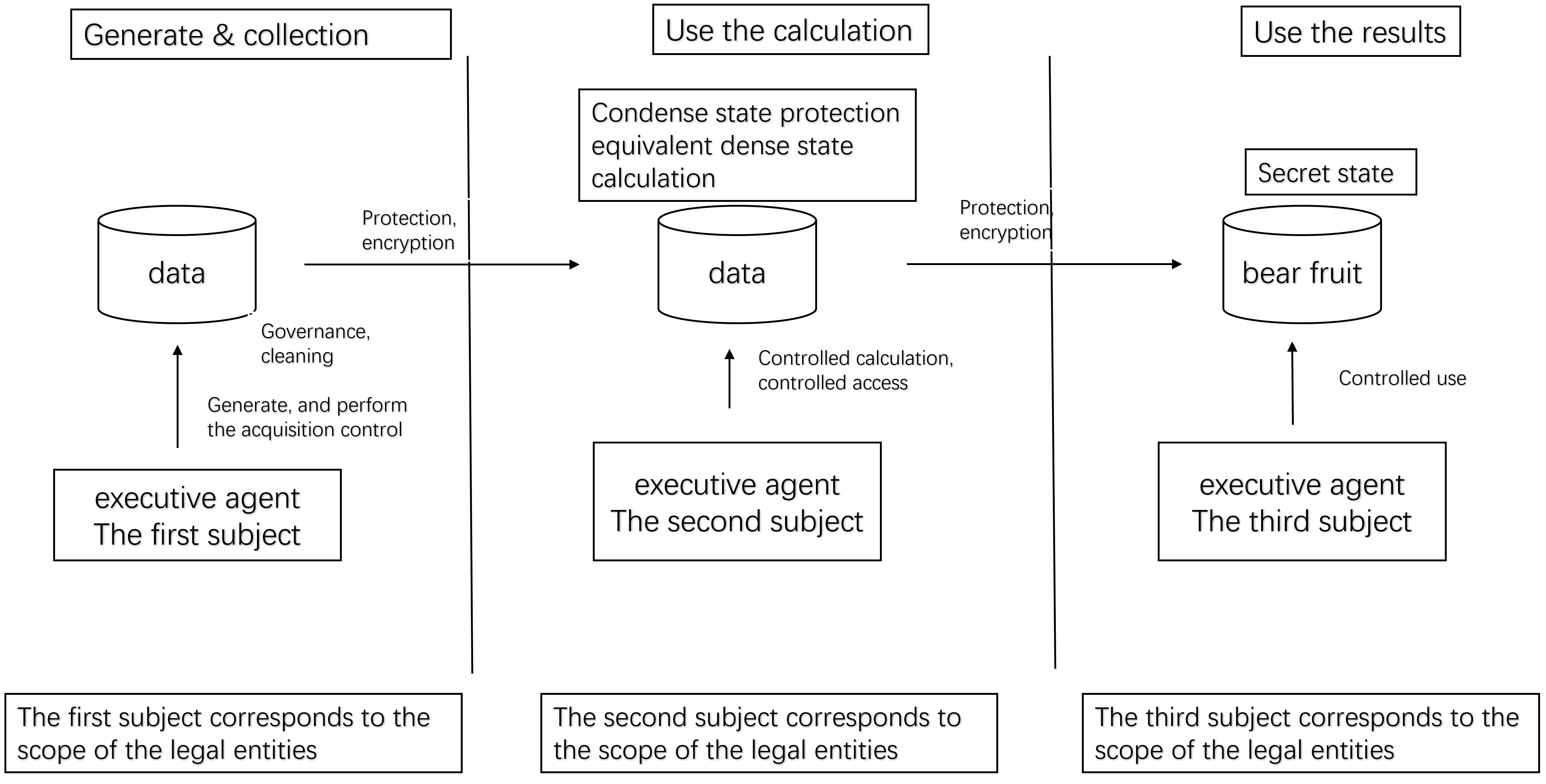}
    \caption{Data Circulation.}
    \label{fig:dataflow}
\end{figure}

The first environmental scenario: when the data are collected and generated, the data are still within the scope of control of the entity (hereinafter referred to as the first entity) who collects and generates the data. The first entity enjoys the rights and undertakes the obligations within the scope of its control, that is, the responsibility of protecting the security of the data. The first entity is responsible for the quality, safety and authenticity of the data.

The second environment scenario: data utilization environment, this time, the executive entity of using data (second entity) and the executive entity of data generation are not unified, in which the following requirements need to be met:
\begin{itemize}
    \item The first executing entity, to the second entity, authorizes partial control of the data, and this partial control, expresses the idea that the second entity is only granted minimal access control to the data for the use of the data. When the use is complete, the first entity is able to withdraw the temporarily authorized access. This process needs to be fully resistant to data vulnerability. It is recommended that it be implemented using the technology at\footnote{Many theories and technical programs do not agree with this point of view, relying on formal laws and regulations contract to ensure data security, from the point of view of this paper is far from enough. Data is complete and substantial protection, to overcome the ``vulnerability'' is the basic requirements of the data circulation, if the circulation process is separated from the first entity outside the plaintext state, that the data has lost the state of being protected. The whole basis of circulation is lost, meaningless. If the technology of ``security computing space for data value'' described in this paper is adopted, it is very convenient to establish a regulatory system and facilitate the determination of the responsibility for data leakage, which can be determined on the basis of the following rules: if the data appear in clear text, it is definitely the responsibility of the first entity.}.

    \item The second entity has control over the acquired data, the use of the data, and the entire process of computation, and has the responsibility for data security (this security responsibility includes storage, transmission, computation, and basic security protection), and it is best to differentiate between the first step of identifying the source of responsibility in the case of a data breach.
\end{itemize}

In addition, data circulation is a multi-stakeholder scenario, and although there is a supervisory party, it does not mean that there is a party that has rights over the other parties, so any entity has the right to supervise based on the use of data provided by itself, even if the data as well as can be accessed and processed by other entities. That is, the first entity has the right to supervise the use of the data provided by the second entity, while retaining the right to process the data provided by itself. This right, of course, must not jeopardize the rights and interests of other parties, and in addition must be guaranteed:
\begin{itemize}
    \item The first entity remains in full control of the data that has been submitted to the second entity.
    \item The second entity cannot use the data provided by the first entity outside of the agreement, even if it is malicious. That is, the second entity can only execute a limited number of algorithms.
\end{itemize}

The third environmental scenario: the second entity hands over the results of the calculation to the executing entity (third entity) that uses the results of the calculation after the data has been used (the calculation is complete).

The third party can use the results according to the agreement, and the third party is responsible for the security of the data (this security includes storage, transmission and computation).

A third entity, even if malicious, cannot enforce a use that violates the covenant.

In addition, all three scenarios also face internal and external security issues, i.e., they all face ``vulnerabilities'' against data, as well as internal and external security issues from the participants.

The same principle applies when more executing entity are involved.

In summary, data circulation occur between multiple entities. It can be summarized that ``data circulation'' exist because there is a need for data to ``cross domains''. This can lead to a series of security and legal issues and challenges when the data is out of the control of the original executing entity. Therefore, in the concept of data circulation in this paper, if there is no change in the legal entity to which the executing entity belongs, it is not a data circulation, but only if there is a change in the executing entity, especially a change in the legal entity to which it belongs, then it is a data circulation.

Therefore, in the event of a change in the executing entity, it is technically and operationally challenging, and legally challenging, to achieve a compliant, reasonable and complete data circulation. In practice, the ``trusted neutral third party'' model is a difficult choice. Trusted neutral third party is an important role in many businesses, let's imagine an all-powerful and good God (although I'm a complete and utter atheist, but the concept of God as a metaphor is easy to understand), this role, he is omnipotent, perfect, he is impartial and selfless without interest tendency, there will be no operational errors, and can resist all the evils, keep any secrets. If every participant, who gives him the data for processing, is safe, that is to say, compliant.

In fact, the current laws and regulations of many countries and regions, the construction of data circulation organizations, or circulation methods, do not understand and apply the privacy-preserving confidential computation technology of ``data availability and invisibility", but only adopt the traditional big data technology, the content of which is to hope that all participants in the processing of the relevant data, working like God. This is only theoretically good. From the analysis of this paper, we can learn that, in practice, such an approach is just to add an impractical implementation of the entity, so the practice is not only high operating costs, management tools, formalized, difficult to regulate, difficult to land the business, the realization of the burden of technical calculations, the heavy burden of security, the risk can not be avoided, and may even introduce new risks. From all perspectives, the technical business program based on a trusted third party is not an optimal program.

\subsection{Three Models of Data Circulation}
In practice, data circulation can be categorized in various ways. One important classification method is based on the type of computational tasks: individual computation, batch computation, and group classification computation. These three models represent typical business scenarios in data circulation, and the associated ethics, regulations, and technologies differ significantly.

The first model: individual computation, refers to the analysis and computation of data for a single individual. In this model, the data analysis and computation process is performed on the data of a single individual to generate specific insights, predictions, or decisions for that individual. This model is widely used in scenarios such as personalized recommendation, personal health diagnosis, and personal credit assessment. Specific examples are given below:

Personal Credit Assessment: Banks use an individual's historical transaction records, loan records, and personal information to analyze his or her creditworthiness in order to decide whether to grant a loan and the amount of the loan.

Personal Health Diagnosis: Hospitals use data from patients' medical records, test results and lifestyle habits to analyze and diagnose their conditions and develop personalized treatment plans.

The second mode: batch computing, is a computational method for unified processing of a group of data or an entire data set. This mode is usually used for data mining, model training, large-scale data analysis, etc. It is suitable for the scenarios that require the extraction of patterns, trends or statistical information from a large amount of data, as well as the more popular large model artificial intelligence training. Specific examples are as follows:

Credit model training: Financial institutions collect credit records and repayment histories of a large number of users and build credit models through data mining techniques for assessing customers' credit risks.

Market Trend Analysis: By analyzing a large amount of data on consumer purchasing behavior and preferences, companies are able to identify market trends and guide product development and marketing strategies.

The third mode: categorical computing, which focuses on dividing the data set into categories or clusters and analyzing each category or cluster with specific computations. This mode is commonly used for cluster analysis, population segmentation, target market analysis, etc., aiming to identify different group characteristics or behavioral patterns from the data. Specific examples are given below:

Market Segmentation: Companies divide the market into segments based on consumers' purchasing history, preferences and demographic characteristics in order to target marketing activities and product positioning.

Health Risk Assessment: Medical research organizations analyze health data of different groups to identify high-risk groups and provide a basis for preventive health interventions.

The difference between the three modes is that, while both batch computing and categorical computing may involve the processing of large amounts of data, batch computing focuses on the overall data set and looks for universally applicable laws and patterns, whereas categorical computing subdivides the data set and then processes and analyzes it separately in order to identify the unique characteristics or needs of each subset. In short, batch computing emphasizes the ``whole'' while classification computing emphasizes the ``differences''. These three models correspond to different regulatory compliance and ethical issues. They have different interest mechanisms, Facing different technical and business challenges, and making different trade-offs.  And require different security measures to deal with data circulation, the detailed discussion of which exceeds the length of this paper. Preliminary analysis of the models described above and the corresponding scenarios shows that there are three main problems in data circulation.

\begin{enumerate}
    \item Cooperation under the malicious model between executing entity, i.e., trust issues
    \item Internal and external data security issues faced by executing entity
    \item Compliance challenges of data circulation, i.e., compliance issue	
\end{enumerate}

Further from business realities, there are further requirements for data circulation, for example:
\begin{itemize}
    \item The ability of the regulator to regulate data circulation before, during, and after the flow of computation, and this ability is not limited to the plain text of the original. Further, the overall environment of data circulation is also included in the scope of regulation.
    \item There exists a credible and operational benefit distribution mechanism where all participants receive a benefit. And it is not one-off but repeated and sustainable.
    \item Participants have one or more corresponding roles, each with clearly defined responsibilities, and when the circulation calculation process is performed, each role performs are involved and their behavior is controlled and audited.
    \item When data (including intermediate data and other forms) and related information are passed between various actors in the process of circulation computing, they are always in a protected state and can resist various malicious attacks. It also guarantees the protection of the key stakeholders of the data, and the entity of data execution has the right and ability to audit its own data.

\end{itemize}

A reasonably complete flow of data should do the following:
\begin{itemize}
    \item Complete supervision, including access certification, approval, calculation process control, result control, and interruptible supervision capability.
    \item Ensure that the interests of all parties are protected and that an adjustable and reliable distribution mechanism can be developed and implemented smoothly.
    \item Clear roles and responsibilities for all parties, clear working interfaces, and tight and seamless chain linkages
    \item Data security throughout the entire process, controlled throughout the process, clear responsibilities, to overcome the ``vulnerability'' of the data, in particular, to be able to realize the destruction of the circulation after the completion of circulation
    \item Data dissemination for potential breaches, with the ability to recognize them, And Appropriate technical measures can be employed to address and prevent these issues.

\end{itemize}

There is already a technical solution that can theoretically meet these requirements, namely, the ``Security Computing Space for Data Value''\footnote{There are some differences with the concepts of data space or safe data space and data base that have existed in the industry for many years. After preliminary analysis, the ``Security Computing Space for Data Value'' has been greatly upgraded and enhanced in terms of compliance, security, and regulation, and the degree of flexibility has been greatly enhanced, and the theoretical foundation is more solid. It can be an upgraded or enhanced version of these concepts.} technical framework, which is mainly realized through privacy-preserving confidential computation as well as hardware and software technologies based on it. The ``Data Value Secure Value Space'' is a complete infrastructure technology framework, which includes the following main aspects:
\begin{itemize}
    \item With hardware, platform, algorithm and data,these four types of computational elements as the regulatory content, cryptography technology combined with privacy and confidentiality calculation is used to realize the comprehensive regulation of data circulation infrastructure.
    \item Through the computational task as the object of supervision, the cryptography technology combined with privacy confidentiality calculation realizes the comprehensive supervision of the circulation process of data circulation.
    \item By combining hardware, platforms, algorithms and data, cryptography techniques are used to form a secure computing space in combination with privacy and confidentiality calculations, and the computing space is authenticated and managed.
    \item Using privacy and confidentiality calculations as the basic technical solution and overcoming the ``vulnerability'' of the data as a means to protect the data throughout the entire process, the data are always encrypted and regulated throughout the process. Whenever there is a change in the executing entity of the data, in particular when the executing entity belongs to a different legal entity, encryption protection is carried out once.

\end{itemize}
For details, refer to the relevant literature.

\section{Conclusion}
In addition to the fact that ``privacy protection'' is a fundamental starting point in data circulation, privacy protection and data protection share some similarities or analogies, One of the common points is the transition from ``confidentiality'' to ``control''. In today's society, a lot of personal information or institutional information is bound to appear in the public domain, or has been disclosed to government agencies and institutions and business organizations, but should not give up protection simply because it has been disclosed. Data circulation is the same, although the data may have been some way to be already open, or has been used, but as data factors, still need a correct, systematic business process system, more systematic infrastructure to support and ensure the full range of safe data circulation.

Based on the analysis of philosophical principles, this paper defines the basic concepts and builds a good Interpretive framework	, which makes the overall structure of data circulation clear and simple, and lays a good foundation for the development of the data industry from both theoretical and practical perspectives.

\theendnotes

\end{document}